\documentclass[a4,12pt]{article}
\usepackage[small,compact]{titlesec}
  
\usepackage[latin1]{inputenc}
\usepackage[T1]{fontenc}
\usepackage{rotating}
\usepackage{multirow}
\usepackage{array}
\usepackage{graphicx}
\usepackage{epstopdf}
 
\usepackage{amsbsy}
\usepackage{amssymb}
\usepackage{amscd}
\usepackage{amsmath}

\usepackage{tabu,import}

 \usepackage{times,epsfig}
 \usepackage{subcaption}
 
\usepackage{cite}
\usepackage{latexsym}
\usepackage{graphicx}
\usepackage{amsmath}
\usepackage{amssymb}
\usepackage{amsfonts}
\usepackage{psfrag}

\usepackage{rotating}

\usepackage{pdflscape}

\newlength{\figwidth}
\newlength{\ffh}
\newlength{\ffw}
\begin{document}

\setlength{\unitlength}{1cm}

\pagestyle{myheadings}

\setcounter{page}{1}

\begin{center}
\large \textsc{Antenna Count for Massive MIMO: 1.9 GHz versus 60 GHz} \\~\\
\normalsize Erik G. Larsson, Link\"oping University, Link\"oping, Sweden\\
\normalsize Thomas L. Marzetta, New York University, NY, USA \\
\normalsize Hien Quoc Ngo,  Queen's University, Belfast, U.K.  \\
\normalsize Hong Yang, Nokia Bell-Labs, Murray Hill, NJ, USA 

\end{center}

\begin{abstract}
If we assume line-of-sight propagation and perfect channel state
information at the base station -- consistent with slow moving
terminals -- then a direct performance comparison between 
Massive MIMO at PCS and mmWave frequency bands is straightforward and
highly illuminating. Line-of-sight propagation is considered favorable
for mmWave because of minimal attenuation, and its facilitation of
hybrid beamforming to reduce the required number of active
transceivers. 

We quantify the number of mmWave ($60$~GHz) service antennas that are
needed to duplicate the performance of a specified number of PCS
($1.9$~GHz) service antennas. As a baseline we consider a modest PCS
deployment of 128 antennas serving 18 terminals. At one extreme, we
find that, to achieve the same per-terminal max-min 95\%-likely
downlink throughput in a single-cell system, $20000$ mmWave antennas
are needed. To match the total antenna area of the PCS array would
require $128000$ half-wavelength mmWave antennas, but a much reduced
number is adequate because the large number of antennas also confers
greater channel orthogonality.  At the other extreme, in a highly
interference-limited multi-cell environment, only $215$ mmWave
antennas are needed; in this case, increasing the transmitted power
yields little improvement in service quality.

\end{abstract}

\section{Introduction}

Massive MIMO \cite{MLYN2016book} and millimeter-wave (mmWave)
communications
\cite{DBLP:journals/pieee/RanganRE14,rappaport2014millimeter,DBLP:journals/access/RappaportSMZAWWSSG13,roh2014millimeter,swindlehurst2014millimeter,andrews2017modeling}
are the two main physical-layer techniques considered for wireless
access in 5G and beyond. The relative merits of one over the other
have been the subject of much public debate, most recently, in the
Globecom 2016 industry panel ``Millimeter wave versus below 5 GHz
Massive MIMO: Which technology can give greater value?''.  The object
of this paper is to give some insight into the relative performances
of the two technologies in a scenario where uncontested models of the
physical propagation channels exist: line-of-sight between the
terminals and the base station(s), and negligible terminal mobility.

\subsection{Massive MIMO Below 5 GHz}

Massive MIMO relies on the use of large antenna arrays at the base
station.  In TDD operation, each base station learns the uplink
channel responses from its served terminals through measurements of
uplink pilots. By virtue of reciprocity of propagation, these
estimated responses are valid also in the downlink. Based on the
estimated channels, the base station performs multiuser MIMO decoding
on uplink and precoding on downlink.  In FDD operation, in contrast,
downlink pilots and subsequent feedback of channel state information
from the terminals are required.

Fundamentally, the permitted mobility is dictated by the
dimensionality of the channel coherence interval -- that is, the
coherence bandwidth in Hertz multiplied by the coherence time in
seconds. In a TDD deployment, the coherence interval determines how
many terminals may be multiplexed, because all terminals must be
assigned mutually orthogonal pilot sequences that fit inside   this
coherence interval.  In an FDD deployment, the coherence interval
dimension limits both the number of base station antennas (as each
antenna needs to transmit a downlink orthogonal pilot) and the amount
of channel state information that each terminal can report.
 
Frequency bands below 5 GHz are particularly attractive for TDD
Massive MIMO, because they support aggressive spatial multiplexing
under high terminal mobility, as proven both by information-theoretic
calculations \cite{MLYN2016book} and by practical experiments
\cite{DBLP:journals/corr/HarrisMVTHLBAE17}. In addition,
below 5 GHz diffraction is significant and electromagnetic waves
penetrate non-metallic objects and foliage well -- enabling the
coverage of wide areas both indoors and outdoors.

\subsection{mmWave Communications}

Millimeter Wave communications typically refers to wireless operations
in the band between 30 and 300 GHz.  Its feasibility has been
demonstrated in practice, for example, using beam-tracking algorithms
in a small-cell environment \cite{qualcomm} and highly directive horn
antennas for transmission in rural areas
\cite{DBLP:conf/mobicom/MacCartneySRXYK16}.  The technology is
considered a contender for the provision of outdoor cellular access
\cite{roh2014millimeter,andrews2017modeling}.

Compared to sub-5 GHz bands, the physics at mmWave bands is different
in several respects:
\begin{enumerate}
\item The effective area of an antenna, for a constant gain pattern,
  scales in proportion to $1/f_c^2$, where $f_c$ is the carrier
  frequency. This means that for a given radiated power, the number of
  receiver antennas required to maintain a predetermined link budget
  scales proportionally to $f_c^2$. Consequently, either numerically
  big arrays, or directive antenna elements, are needed to facilitate
  transmission over large distances.

\item mmWaves do not penetrate solid objects well, and propagation is
  dominated by the presence of unobstructed direct paths and specular
  reflection.  Thus, the conditions favorable to mmWave entail either
  line-of-sight, or a few reflections.
  
\item The Doppler frequency is proportional to the carrier frequency,
  implying increasing difficulty in acquiring channel state
  information for high mobility terminals.

  Fortunately, as the propagation becomes increasingly line-of-sight,
  the acquisition of complete channel responses, for the purpose of
  precoding and decoding, is not necessary. The identification of a
  line-of-sight path, potentially combined with a few reflected paths,
  is sufficient in order to obtain channel estimates that are
  sufficiently accurate.
\end{enumerate}

It may be questioned whether spatial multiplexing to a large number of
users is relevant for mmWave communications, since such systems
usually cover relatively small cells. However, future applications may
very well require the service of much greater densities of terminals
at much higher data rates as compared to today.  Examples that are
discussed currently include virtual and augmented reality, and
transmission of uncompressed real-time imagery and video (for machine
learning purposes) for example between vehicles and infrastructure
(V2X).

Indeed, it is frequently stated that mmWave is favorable for Massive
MIMO multiplexing because the small size of each antenna permits the
deployment of numerically large arrays in a small space. This argument
has been accepted uncritically, in the absence of analyses regarding
how many antennas would actually be required.

\subsection{The Role of Bandwidth}

An argument in favor of mmWave communications is the
availability of huge bandwidths. While this is certainly correct,
extreme bandwidth is only useful if there is an available reserve of
effective power.  Specifically, by the Shannon-Hartley formula,
capacity scales as $B\log_2(1+P/(BN_0))$, where $B$ is the bandwidth
in Hertz, $P$ is the received power in Watt, and $N_0$ is the noise
spectral density in Watt/Hertz.  The capacity is an increasing
function of $B$, for fixed $P$.  However, the returns of increasing
$B$ are diminishing: if $B\to\infty$ for fixed $P$, then the capacity
approaches a \emph{finite} limit equal to $(P/N_0 ) \log_2(e) $.

As an example, suppose that nominally 10 Watt of radiated power over a
bandwidth of 20 MHz is required to sustain a rate of 60 Mbps. If the
bandwidth is increased 50 times, to 1 GHz, then in order to also
achieve a 50-fold increase in capacity, the radiated power must be
increased 50 times, to 500 Watt. If a smaller capacity increase can be
accepted, less power is required.  For example, suppose that with a
50-fold increase in bandwidth, a mere 25-fold increase in rate is
satisfactory. Then the required radiated power over 1 GHz bandwidth is
only 131 Watt.
 
The attractiveness of extreme bandwidth is further diminished because
it reduces the signal-to-noise ratio (SNR) of received pilot signals,
and consequently degrades the quality of the channel estimates.  Let
$\rho_0=P/(B_0N_0)$ be the SNR that occurs for some reference
bandwidth $B_0$. Then the throughput (for sufficiently high bandwidth)
is approximately $B \log_2(1 + \rho_0 B_0^2/B^2)$, so at some point,
additional bandwidth results in smaller throughput. This holds
irrespective of whether the bandwidth is used to serve one terminal,
or whether several terminals are orthogonally multiplexed in different
parts of the band.  This ``squaring effect'' that results when channel
estimates are obtained from uplink pilots is explained in more detail
in \cite[Chap.~3]{MLYN2016book}.  Only if the channel estimates are
obtained from a parametric model (e.g., estimation of
direction-of-arrivals), then the throughput scaling with $B$ can be
more favorable.

\section{Sub-5 GHz versus mmWave Comparison}

To maintain a constant power-aperture product with increasing carrier
frequency would require the number of antennas to grow with the square
of the carrier frequency. However, the increasing number of antennas
facilitates spatial multiplexing because of greater channel
orthogonality, so some reduction in the power-aperture product is
permissible.  Specifically, the probability that two channel responses
are similar decreases sharply when increasing the number of antennas
in the array. This results in a new, intriguing tradeoff, if the
multiplexing of many simultaneous terminals is of interest, and the
provision of uniformly good quality of service in the cell is of
concern. Nominally, considering only the path loss, $(60/1.9)^2\approx
1000$ times more antennas would be required at mmWave (60 GHz) as
compared PCS (1.9 GHz) to maintain the same link budget -- implying,
that 128000 mmWave antennas would be required to match 128-antenna
PCS.  While this is an important conclusion per se, there is also a
countervailing effect: in a scenario where many terminals are
spatially multiplexed, the PCS system may be at a disadvantage because
its smaller number of antennas renders it much more likely that some
terminal must be dropped from service due to channel
non-orthogonality. Thus, the proportionate number of extra antennas
required for the mmWave carrier may be considerably smaller than
$1000$.  This effect is particularly pronounced under line-of-sight
conditions \cite[Section~7.2.3]{MLYN2016book}.

\begin{figure}[t!]
\begin{center}
      \begin{psfrags}
        \psfrag{lam}[Bc][Bl][1.0]{$\lambda/2$}
        \psfrag{Mantenna}[Bc][Bl][1.0]{$M$-antenna}
        \psfrag{bs}[Bc][Bl][1.0]{base station}
\includegraphics[width=15cm]{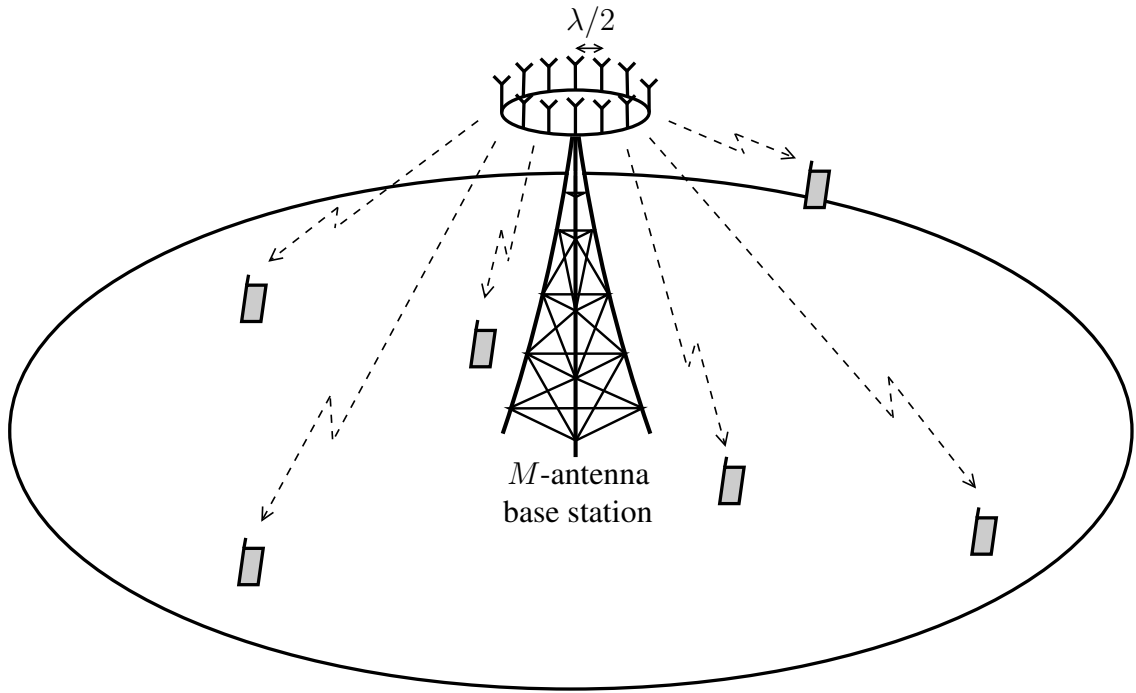}
      \end{psfrags}
\end{center}
\caption{Single cell with circular base station array and
  line-of-sight to all terminals. The base station equipped with $M$
  antennas (half-wavelength spacing) simultaneously serves multiple
  single-antenna terminals. All terminals are located uniformly at
  random inside the cell. \label{fig:cell}}
\end{figure}

The following discussion will quantitatively address this point, and
specifically the question: For a predetermined quality-of-service, how
many more antennas will a mmWave system require, compared to the PCS
system?

\subsection{Single Cell}

We first consider a single Massive MIMO cell, in which a circular base
station array, comprising $M$ antennas at half-wavelength
($\lambda/2$) spacing (Figure~\ref{fig:cell}) serves a multiplicity of
terminals, randomly located in the cell.  The array diameter
(aperture) is $d=M\lambda/(2\pi)$.  A number of terminals are placed
uniformly at random inside the cell, and multiplexed simultaneously in
the same time-frequency resource.

There is line-of-sight between the base station antennas and the
terminals.  The assumption of LoS propagation is, in fact, realistic
for some scenarios (large crowds of people in a stadium, for example,
or in Times Square, New York). Non-line-of-sight operation would hurt
mmWave more than PCS. In addition, a rich scattering environment would
further hamper the use of hybrid (part analog, part digital)
beamforming schemes for mmWaves -- the only hope for reducing the huge
number of active antenna circuits.  Note also that in line-of-sight,
Friis' free-space propagation equation holds exactly, facilitating
analysis without reliance on empirical path loss models.

There is negligible mobility. This enables the base station to learn
the channel responses with arbitrarily good accuracy, and it makes
outage capacity a legitimate performance measure. This assumption
also, substantially, eliminates the distinction between TDD and FDD
mode operation.

We focus on this scenario, because of its simplicity and ease of
analysis, and because, if anything, it is biased in favor of
mmWave. Any departure from line-of-sight propagation would create more
difficulties for mmWave than for PCS. Likewise, any terminal mobility
would require the expenditure of significant resources on channel
estimation, to the greater detriment of mmWave because of its 30-times
larger Doppler frequency. Notwithstanding, this may be a realistic
model for, again, operation in a stadium or in Times Square.

The base station performs zero-forcing (ZF) decoding and precoding,
which is known to be more effective than maximum-ratio (MR) decoding
and precoding for obtaining effective signal-to-noise-and-interference
ratios (SINRs) significantly greater than one
\cite{MLYN2016book}. Also, under line-of-sight propagation, there is a
significant chance for high non-orthogonality between channels, which
is precisely where ZF is effective \cite{hong-preprint}. Efficient
hardware implementations have been demonstrated for reasonably-sized
systems \cite{Prabhu2017}. Max-min power control ensures that every
terminal enjoys the same effective SINR
\cite{MLYN2016book,hong-preprint}.

The random location of each terminal induces a random
path-loss. Standard Massive MIMO analysis techniques yield an SINR --
and an accompanying lower bound on the achievable rate -- for each
terminal, in terms of the path loss and the power control
coefficients. In turn, the power control coefficients are chosen
according to the max-min criterion, as described in
\cite[Chap.~5]{MLYN2016book} and \cite{hong-preprint}. A repetition of
the procedure for many independent realizations of terminal location
yields an empirical cumulative distribution for the max-min SINR.

Table~\ref{tab:1} quantitatively summarizes all relevant parameters
used in this case study.  We adopt isotropic antenna gains of $0$~dBi
solely for simplicity. A more realistic model would, of necessity,
have directional dependence which would introduce additional random
components to the propagation, e.g., the orientation of the terminal
antenna. Owing to the no-mobility assumption, the choice of duplexing
(TDD or FDD) is immaterial.

\begin{table}[t!]
\centerline{
  \tabulinesep=1mm  \taburulecolor{gray}
 \begin{tabu}{|c|c|}
   \hline
   Number of simultaneously multiplexed terminals & $18$ \\
 Distribution of terminal locations  & uniform within circular cell  \\
  Cell radius & 250 m \\
 Base station height over ground & 30 m \\
 Propagation model & line-of-sight (Friis free-space equation) \\
 Base station antenna element gain & 0~dBi \\
 Terminal antenna gain & 0~dBi \\
  Power control & max-min SINR fairness (uniform quality-of-service) \\
 Bandwidth & 50 MHz \\
 Terminal height over ground & 1.5 m \\
 Base station radiated power (downlink) & 2 W \\
 Terminal radiated power (uplink) & 200 mW \\
 Base station noise figure & 9 dB \\
 Terminal noise figure & 9 dB \\
\hline
 \end{tabu}
}
\caption{\label{tab:1}Massive MIMO system parameters.}
\end{table}

\begin{figure}[t!]
\centering
\subimport{CDFs/}{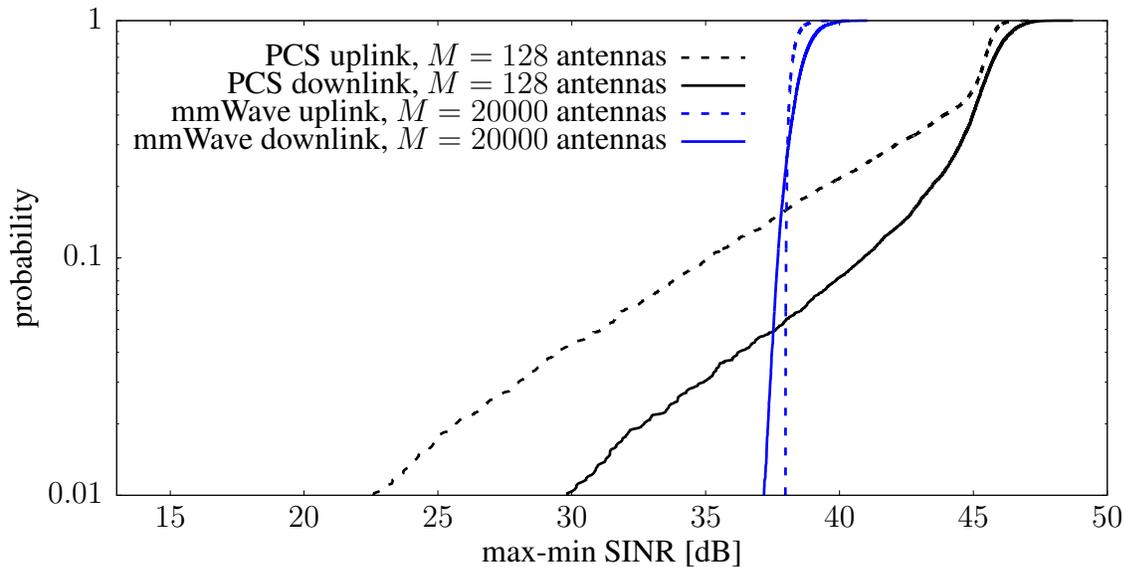} 
\caption{\label{fig:1}Uplink and downlink SINR in the PCS  ($f_c=1.9$~GHz) respectively
 mmWave  ($f_c=60$~GHz) systems, for a single-cell deployment.}
\end{figure}

Figure~\ref{fig:1} shows the cumulative probability distribution of
the effective SINR on uplink and downlink, for the PCS and the mmWave
systems. All randomness originates from the random terminal
locations. The PCS system has 128 antennas and the mmWave system has
20000 antennas.  In Figure~\ref{fig:1}, the 95\%-likely downlink SINR
(the point at cumulative probability 0.05 in the plots --
approximately $38$~dB) is comparable for both systems. The mmWave
system has 156 times more antennas than the PCS system. This is
substantially less than predicted by a path-loss analysis
($(60/1.9)^2\approx 1000$ times). The reason is the improved
orthogonality conferred by the larger number of antennas at
mmWaves. This improved orthogonality is also reflected by the
increased steepness of the curves.

Table~\ref{tab:2}(a) shows the required number of base station antennas,
and the corresponding array diameter, for different 95\%-likely uplink
SINRs. The mmWave array comprises a fairly large number of antennas
but is geometrically rather compact. On downlink,  fewer
antennas are required in either case (not shown in the table).

\begin{figure}[t!]
\centerline{   \subimport{CDFs/}{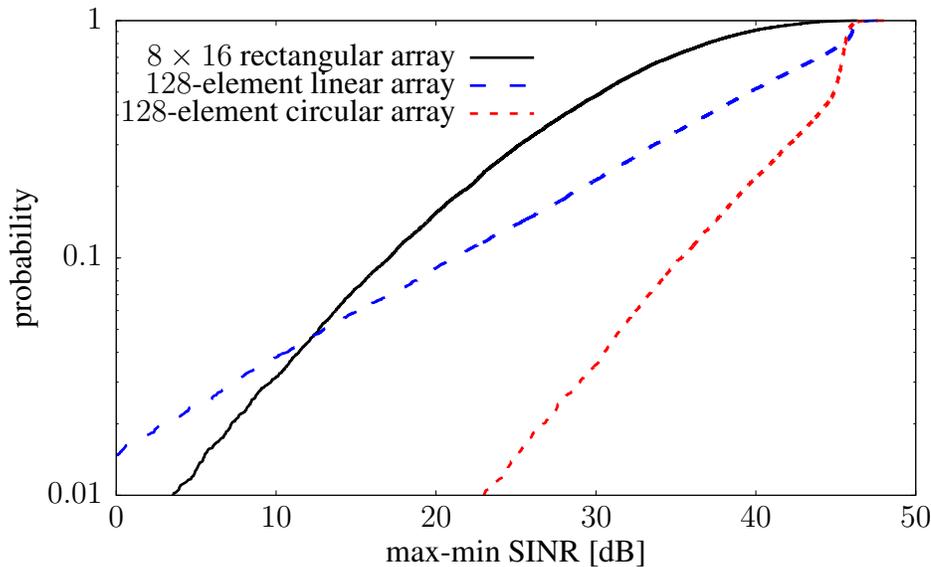} }
\caption{\label{fig:geo}Uplink SINRs for different array geometries, in the single-cell case
and for the PCS system.
In the rectangular and linear arrays, the distance between neighboring elements is $\lambda/2$.
In the circular array, the arc distance between neighboring antennas is $\lambda/2$.}
\end{figure}

The following remarks are in order:
\begin{enumerate}
\item The 95\%-likely SINR targets in Table~\ref{tab:2}(a) range from
  moderate to very high.  From a purely information-theoretic point of
  view, for a given SINR, $\log_2(1+\mbox{SINR})$ bits/s/Hz could be
  achieved. However, for the high SINR values, the resulting
  modulation scheme would require large constellation sizes and
  therefore may be sensitive to phase noise.

\item There is no interference, neither within the cell (owing to the
  perfect-CSI assumption), nor from other cells (owing to the
  single-cell assumption).

\item Hypothetically, if all channels were orthogonal and all
  terminals were subject to the same path loss, the uplink-downlink
  SINR difference would be equal to the power imbalance between uplink
  and downlink, $(18\times 0.2) /2=1.8 \approx 2.6$~dB. Considering
  the path loss differences and given that we use max-min power
  control, most of the terminals do not expend full power in the
  uplink. However, full power is expended and shared among all
  terminals in the downlink. So the power imbalance between uplink and
  downlink will be smaller than $2.6$~dB. This is evident in
  Figure~\ref{fig:1} (the mmWave curves) where the channels are nearly
  orthogonal for large numbers of service antennas.

With far fewer antennas, the PCS channels are far less orthogonal. The
advantage of power pooling in the downlink is seen in Figure~\ref{fig:1} (the PCS curves).
  
\item The quantitative results are somewhat pessimistic since the
  max-min fairness power control criterion forces the simultaneous
  service of every terminal in the cell with the same SINR. If two
  terminals are located close to one another (with nearly the same
  direction-of-arrivals), then the resulting channel is
  ill-conditioned and much power needs be expended to invert
  it. Better performance could be obtained by scheduling one of those
  terminals on an orthogonal resource (for example, a different
  subcarrier), or a much reduced level of service.
  
  However, numerical experiments not disclosed here have shown that
  this effect is rather minor. Also, importantly, the effect is much
  larger for small numbers of base station antennas. Hence, insofar a
  required-number-of-antennas comparison is concerned, the inclusion
  of an orthogonal scheduling option would favor the PCS system much
  more than the mmWave system.

\item The use of antennas with larger physical aperture (i.e.,
  intrinsic directivity) would not change the conclusions, because
  such use of directional antennas is essentially equivalent to
  sectorization. The problem is that to exploit the intrinsic gain,
  the antennas must \emph{a priori} point into a given
  direction. Consequently, in the array, only a subset of the antennas
  will be useful when serving a particular terminal. This negatively
  impacts both the resulting channel gain (in terms of reduced
  effective aperture, since only few of the antennas point into the
  direction of a randomly selected terminal) and -- importantly -- the
  channel orthogonality (see, e.g, Chapter 7 in \cite{MLYN2016book}).

\item The array geometry can have a large impact on the performance.
  Figure~\ref{fig:geo} illustrates this point quantitatively, by
  comparing a rectangular, a linear, and a circular array for the
  uplink. The circular array significantly outperforms the other two.
  The reason is that neither the rectangular nor the linear array has
  any resolution capability in its endfire directions.  Similar
  observations hold for the downlink (not shown in Figure~\ref{fig:geo}).

\item The mmWave system may utilize multiple antennas more readily at
  the terminals, resulting in a received power gain, as well as
  interference nulling capabilities. This is possible also in the PCS
  system, but antenna arrays at the terminals will be bulkier.
  
\item The mmWave system may have access to much larger bandwidths,
  although correspondingly higher radiated powers are required to
  efficiently exploit these bandwidths -- as discussed above.

\item Achieving high SNR requires array diameters on the order of 1
  to 3 meters.  While this physical size is substantial, nothing
  prevents such antenna arrays from being engineered. Traditional
  arrays could be constructed using metallic frames, or antennas could
  be mounted on pre-existing buildings, concrete structures or steel
  frames. Yet, we speculate that more advantageously, structures may
  be built from modern materials (e.g., carbon-fiber reinforced
  polymer) with physically small antenna elements, cables and all
  required electronics integrated inside.
 
\item The mmWave array is certainly smaller than the PCS array, but
  the maximum ratio of diameters is  far
  smaller than the wavelength ratio of $32$.

\item Some savings in mmWave electronics may be realized by employing
  hybrid beamforming that exploits the structure of line-of-sight
  propagation \cite{DBLP:journals/twc/AyachRAPH14}.  Yet, the angle
  spread associated with non-line-of-sight propagation will limit the
  extent to which hybrid beamforming may be used.

\item Non-line-of-sight propagation would cause signals to attenuate
  at a rate much greater than in free space, with mmWave more
  disadvantaged than PCS.  Our line-of-sight analysis ignores the
  sensitivity of mmWave signals to blockage. Blockage by the human
  body alone can yield tens of dB of losses
  \cite{collonge2004influence}. A possible remedy to that problem is
  to establish coverage from multiple access points.

\item The zero-mobility assumption, and the consequent unlimited
  coherence time, makes our scenario very favorable for mmWave
  technology.  In reality, channel estimation and/or beam training
  will be required.  These tasks are more challenging for mmWave
  technology, since the channels change at thirty times the rate of
  PCS channels.
\end{enumerate}

\begin{table}[t!]
\centering
\begin{subtable}{1.0\linewidth}
\centering
  \tabulinesep=1mm  \taburulecolor{gray}
 \begin{tabu}{c|c |c| c|c|}
     \cline{2-5}
 &  \multicolumn{2}{|c|}{Required number} & \multicolumn{2}{|c|}{Array diameter} \\
 &  \multicolumn{2}{|c|}{of antennas ($M$)} & \multicolumn{2}{|c|}{ (meter)} \\
 \hline
\multicolumn{1}{|c|}{ 95\%-likely max-min SINR}        &   \textbf{PCS}      &   \textbf{mmWave} &  \textbf{PCS}        &   \textbf{mmWave} \\
\multicolumn{1}{|c|}{}    &   ($1.9$~GHz)        &    ($60$~GHz) &   ($1.9$~GHz)        &    ($60$~GHz)  \\
 \hline
\multicolumn{1}{|c|}{ $5$~dB}        &   $33$   & $160$ & $0.83$ & $0.13$ \\
\multicolumn{1}{|c|}{ $10$~dB}  &  $40$ & $250$ & $1.0$ & $0.20$ \\
\multicolumn{1}{|c|}{ $15$~dB} & $54$  & $360$ & $1.4$ & $0.29$ \\
\multicolumn{1}{|c|}{ $20$~dB} & $64$ & $560$ & $1.6$ & $0.45$ \\
\multicolumn{1}{|c|}{ $25$~dB} & $90$ & $1100$ & $2.3$ & $0.87$ \\
\multicolumn{1}{|c|}{ $30$~dB} & $110$ & $4000$ & $2.8$ & $3.2$ 
 \\ \hline
 \end{tabu}
\caption{Number of antennas, and resulting diameter of
  the circular array, required to obtain a given 95\%-likely uplink
  SINR, for the single-cell setup.}
  \label{tab:2a}
\end{subtable}%

~

~

~

\begin{subtable}{1.0\linewidth}\centering 
  \tabulinesep=1mm  \taburulecolor{gray}
 \begin{tabu}{c|c |c| }
     \cline{2-3}
 &  \multicolumn{1}{|c|}{\textbf{PCS}} & \multicolumn{1}{|c|}{\textbf{mmWave}} \\ 
 \hline
    \multicolumn{1}{|c|}{ spectrum available for 5G}        &   some new & much new \\
 \hline
    \multicolumn{1}{|c|}{ market value of spectrum}        &   U.S.\ operators paid  \$40 billion & not known  \\
    \multicolumn{1}{|c|}{ }        &    for 65 MHz  in 2014 &  \\
 \hline
\multicolumn{1}{|c|}{ required $M$}        &   some 100  & some 10000 \\
 \hline
   \multicolumn{1}{|c|}{ uniform quality-of-service}        &   feasible within a cell,  & feasible within a cell, \\
   \multicolumn{1}{|c|}{ (max-min power control)}        &  regardless of propagation \cite{MLYN2016book}   & if line-of-sight  (this paper) \\
 \hline
 \multicolumn{1}{|c|}{ probability that two channels}        &   small    & very small \\
 \multicolumn{1}{|c|}{  are considerably correlated}        &       &  \\
 \hline
 \multicolumn{1}{|c|}{ mobility}        &   handles well & very challenging \\
 \multicolumn{1}{|c|}{ }        &    & (30$\times$ Doppler) \\ 
 \hline
    \multicolumn{1}{|c|}{ blocking (e.g.\ by human body)}        &   not a problem & serious issue \\
 \hline
    \multicolumn{1}{|c|}{ nature of interference }        &   stationary & non-stationary \\
    \multicolumn{1}{|c|}{ (in environment with mobility) }        &    &  \\
 \hline
 \end{tabu}
\caption{Similarities and differences between PCS
and mmWave Massive MIMO.}
\label{tab:2b}
\end{subtable}
\caption{\label{tab:2}Comparison between Massive MIMO at PCS and mmWave frequency bands.}
\end{table}

Table~\ref{tab:2}(b) summarizes similarities and differences between PCS
and mmWave Massive MIMO.

\subsection{Multiple Cells}

A legitimate question is whether the above single-cell example is too
``extreme'' and to what extent the conclusions would change in a
multi-cell environment. Naturally, the salient features of PCS versus
mmWave operation are at work fully in the single-cell scenario.
However, in a multi-cell system, intercell interference may become
significant and eventually the main limiting factor.
  
A proper multi-cellular analysis would require many additional
assumption to be made (all subject to controversy), including the
selection of appropriate terminal-to-base station assignment
algorithms and power control schemes. In particular, while max-min
fairness power control is feasible \emph{within a cell}, equalization
of the throughputs \emph{over multiple cells} is not scalable, so
heuristic schemes are required that achieve fairness between the
different cells. While these issues are rather well understood for PCS
systems under an independent Rayleigh fading assumption (see, e.g.,
\cite[Chaps.~5--6]{MLYN2016book}), they are not well investigated for
mmWave systems. In particular, modeling of intercell interference is
non-trivial at mmWaves, first, because that interference is likely to
be non-stationary, and second, because even though propagation within
the cell is line-of-sight, interference from other cells may not be.

\begin{figure}[t!]
\centering
\subimport{CDFs/}{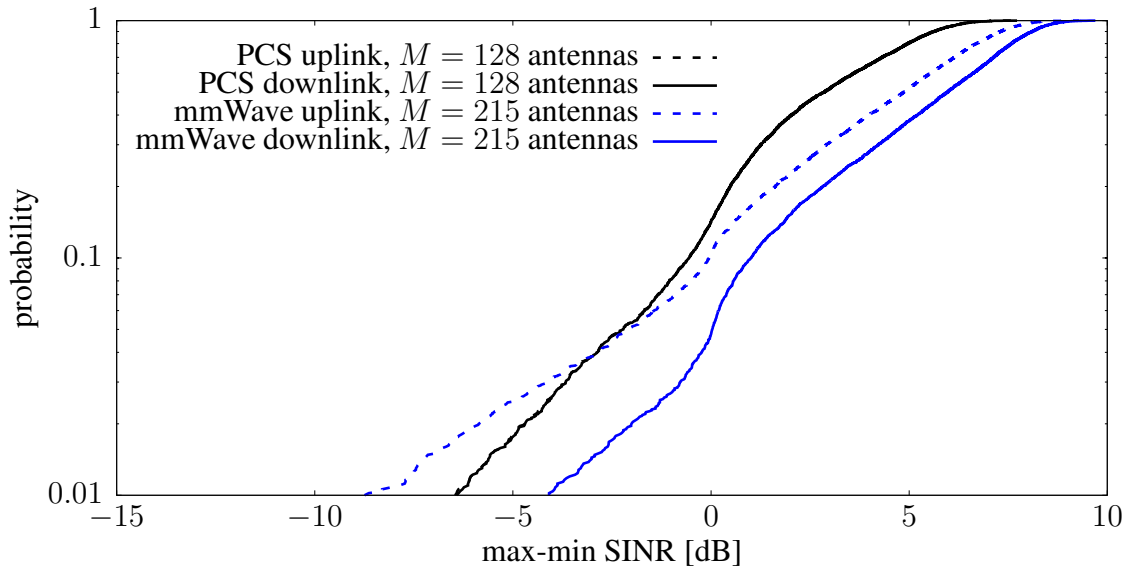} 
\caption{\label{fig:1b}Uplink and downlink SINR in the PCS respectively
  mmWave systems, for a multi-cell deployment with line-of-sight between all base stations
  and all terminals,  no blocking, and system-wide max-min fairness power control.
The uplink and downlink curves for PCS overlap in the plot.}
\end{figure}

Notwithstanding, the effects of intercell interference can be
illustrated through simple modeling. To examine the effect to its
extreme, we consider a system with seven neighboring circular cells,
using system-wide max-min fairness power control, and assume that
there is line-of-sight propagation between all base stations and
terminals.  This results in an extremely interference-limited
scenario. Figure~\ref{fig:1b} shows the corresponding results. In this
setup, a 128-antenna PCS system offers roughly equivalent performance
to a 215-antenna mmWave system.  Due to the high amount of intercell
interference, here the power advantage of PCS becomes relatively
unimportant. This is why much fewer mmWave antennas are required to
make up the power disadvantage.
 
Note that under the adopted negligible-mobility model, perfect
channel state information is available at the base station and then
interference on the pilot channels becomes a non-issue. Consequently,
in a multi-cell environment only non-coherent interference is
present, whose magnitude is independent of the number of base station
antennas.

\section{Discussion and Conclusions}

The main message is that when increasing the number of base station
antennas, the channel orthogonality improves.  While a link budget
analysis would predict that a factor $(f_2/f_1)^2$ more antennas are
required when going from a carrier frequency $f_1$ to a carrier
frequency $f_2$, in reality that number can be much smaller.  While
previous work exists \cite{heath-ctw} that has addressed the question
of relative performance between different carrier frequencies, the
point on channel orthogonality -- which proves to be exceedingly
important in context -- has not been articulated elsewhere to our
knowledge.

Notwithstanding, in a noise-limited environment (the single-cell
example), the number of antennas required to compensate for a
decreased effective antenna area can be quite large. Specifically, for
a particularly simple scenario -- line-of-sight propagation and
perfect CSI -- we compared 128-antenna PCS (1.9 GHz) massive MIMO with
mmWave (60 GHz):
 \begin{itemize}
 \item For a noise-limited single-cell setup, a link budget
   calculation might predict that $128\times (60/1.9)^2 \approx
   128000$ mmWave antennas are needed to match the performance of
   128-antenna PCS MIMO, but there is a countervailing effect in that
   increasing the number of antennas improves channel orthogonality so
   that only 20000 antennas are required.
\item In contrast, in a severely interference-limited multi-cell
  scenario, the numbers of antennas required at the two different
  carrier frequencies are comparable (128 versus 215). In this
  interference-limited regime, the power advantage of PCS becomes
  unimportant and the performance becomes independent of the carrier
  frequency.
\end{itemize}

\section{Acknowledgements}
The authors thank the Swedish research council (VR) and ELLIIT for funding parts of this work.

\bibliographystyle{IEEEtran}

\section{Biographies}

Erik G. Larsson is Professor at Link\"oping University, Sweden. He
co-authored \emph{Fundamentals of Massive MIMO} (Cambridge, 2016) and
\emph{Space-Time Block Coding for Wireless Communications} (Cambridge,
2003). Recent service includes membership of the IEEE Signal
Processing Society Awards Board (2017--2019), and the \emph{IEEE
  Signal Processing Magazine} editorial board (2018--2020). He
received the {IEEE Signal Processing Magazine} Best Column Award
twice, in 2012 and 2014, the IEEE ComSoc Stephen O. Rice Prize in
Communications Theory 2015, the IEEE ComSoc Leonard G. Abraham Prize
2017 and the IEEE ComSoc Best Tutorial Paper Award 2018. He is a
Fellow of the IEEE.

Thomas L. Marzetta is Distinguished Industry Professor in the ECE
Department at NYU Tandon School of Engineering. He joined NYU in 2017
after spending thirty-nine years in three industries: petroleum
exploration, defense, and telecommunications. He is the originator of
Massive MIMO. Recognition for his achievements includes the IEEE
Communications Society Industrial Innovation Award, the IEEE Stephen
O. Rice Prize, the IEEE W. R. G. Baker Award, and the Honorary
Doctorate from Link\"oping University.

Hien Quoc Ngo received his Ph.D. in 2015 from Link\"oping University,
Sweden. He is now an assistant professor at Queen's University
Belfast, U.K. His research interests are the applications of
mathematical, statistical, random matrix, optimization theories, and
signal processing to wireless communications. He received the IEEE
ComSoc Stephen O. Rice Prize in 2015, the IEEE ComSoc Leonard
G. Abraham Prize in 2017, and the Best Ph.D. Award from EURASIP in 2018.

Hong Yang received the Ph.D. degree in applied mathematics from
Princeton University, USA. He is a member of technical staff with the
Mathematics of Networks and Communications Research Department, Nokia
Bell Labs, Murray Hill, NJ, USA, where he conducts research in
communications networks. He has coauthored many research papers in
wireless communications, applied mathematics, control theory, and
financial economics. He coinvented many U.S. and international
patents. He coauthored the book \emph{Fundamentals of Massive MIMO}.

\end{document}